\documentclass[twocolumn,prb,showpacs]{revtex4}
\usepackage{graphics}
\usepackage{graphicx}
\usepackage{amsmath}
\usepackage{amssymb}
\usepackage{subfigure}
\usepackage{longtable}
\usepackage{multirow}
\usepackage{bm}
\usepackage{hyperref}

\newcommand{\ve}[1]{\bm{\mathrm{#1}}}
\newcommand{\rr}{\mathbf{r}}
\newcommand{\rp}{\mathbf{r}^{\prime}}
\newcommand{\np}{n^{\prime}}
\newcommand{\w}{\omega}
\newcommand{\kk}{\mathbf{k}}
\newcommand{\kq}{\mathbf{k} + \mathbf{q}}

\begin{document}
\title{Linear density response function in the projector-augmented wave method:
Applications to solids, surfaces, and interfaces}

\author{Jun Yan}
\affiliation{Center for Atomic-scale Materials Design, Department of
Physics \\ Technical University of Denmark, DK - 2800 Kgs. Lyngby, Denmark}
\author{Jens. J. Mortensen}
\affiliation{Center for Atomic-scale Materials Design, Department of
Physics \\ Technical University of Denmark, DK - 2800 Kgs. Lyngby, Denmark}
\author{Karsten W. Jacobsen}
\affiliation{Center for Atomic-scale Materials Design, Department of
Physics \\ Technical University of Denmark, DK - 2800 Kgs. Lyngby,
Denmark}
\author{Kristian S. Thygesen}
\email{thygesen@fysik.dtu.dk}
\affiliation{Center for Atomic-scale Materials Design, Department of
Physics \\ Technical University of Denmark, DK - 2800 Kgs. Lyngby, Denmark}

\date{\today}

\begin{abstract}
  We present an implementation of the linear density response function
  within the projector-augmented wave (PAW) method with applications
  to the linear optical and dielectric properties of both solids,
  surfaces, and interfaces. The response function is represented in
  plane waves while the single-particle eigenstates can be expanded on
  a real space grid or in atomic orbital basis for increased
  efficiency. The exchange-correlation kernel is treated at the level
  of the adiabatic local density approximation (ALDA) and crystal
  local field effects are included. The calculated static and
  dynamical dielectric functions of Si, C, SiC, AlP and GaAs compare
  well with previous calculations.  While optical properties of
  semiconductors, in particular excitonic effects, are generally not
  well described by ALDA, we obtain excellent agreement with
  experiments for the surface loss function of the Mg(0001) surface
  with plasmon energies deviating by less than 0.2 eV. Finally, we
  apply the method to study the influence of substrates on the plasmon
  excitations in graphene. On SiC(0001), the long wavelength $\pi$
  plasmons are significantly damped although their energies remain
  almost unaltered. On Al(111) the $\pi$ plasmon is completely
  quenched due to the coupling to the metal surface plasmon.
\end{abstract}

\maketitle

\section{Introduction}
Time-dependent density functional theory (TDDFT)\cite{Gross_L84} has
been widely used to calculate optical excitations in molecules and
clusters as well as the optical and electron energy loss spectra of
bulk semiconductors, metals and their surfaces\cite{TDDFT_review}.
The excitation energies and oscillator strengths of both
single-particle and collective electronic excitations are determined
by the frequency-dependent linear density response function $\chi(\ve
r, \rp, \w)$ giving the density response at point $\ve r$ to first order in a time-dependent perturbation of frequency $\omega$ applied at point $\rp$,
\begin{equation}
\delta n(\ve r, \w) = \int
d\ve r\chi(\ve r, \rp, \w) \delta V_{\mathrm{ext}}(\rp, \w).
\end{equation}
For finite systems, $\chi$ can be efficiently
calculated by inverting an effective Hamiltonian in the space of particle-hole
transitions. For the
practically relevant case of frequency-independent
exchange-correlation kernels this formulation leads to the well known
Casida equation\cite{Casida_98}. For extended systems, it is more
convinient to express $\chi$ in a basis of plane
waves\cite{Adler_B62,Wiser_B63,Louie_B87} where it has the generic
form $\chi_{\ve G \ve G^{\prime}}(\ve q, \w)$, with $\ve G$ being 
reciprocal lattice vectors and $\ve q$ being wavevectors in the first
Brillouin zone (BZ). 

In this paper we focus on the electronic response function of extended
systems treating electron-electron interactions at the level of the
random phase approximation (RPA) and the adiabatic local density
approximation (ALDA). For many extended systems such a description is
insufficient to account for optical excitations because the electron-hole attraction
is not properly accounted for.
However, dielectric properties, in particular collective plasmon
excitations, are generally accurately reproduced by this
approach\cite{Echenique_RPP07,Rocca_95}, and quantitative agreement
with electron energy loss experiments have been reported for bulk
metals\cite{Eguiluz_L93,Onida_B02}, surfaces\cite{Tsuei_SS91,Echenique_N07},
graphene-based
systems\cite{Rubio_L02,Reining_L08}, 
semiconductors\cite{Hanke_B00,Reining_L01} and even
supercondutors\cite{Eguiluz_L02}. Furthermore, the accurate evaluation of the
density response function at the RPA or ALDA level is a prerequisite for
implementation of most
post-DFT schemes, such as RPA correlation energy\cite{Kresse_RPA},
exact-exchange optimized-effective-potential
methods\cite{Gorling_L97}, the GW approximation for quasi-particle
excitations\cite{Louie_B86,Rubio_RMP02}, and the Bethe-Salpeter
equation\cite{BSE_51,Rubio_RMP02} for optical excitations.

Here we present an implementation of the density response function
within the electronic structure code \textsc{gpaw}\cite{GPAW_B05,GPAW_10} which
is based on the projector augmented wave (PAW)
methodology\cite{Blochl_B94,Blochl_03} and represents wave functions
on real space grids or in terms of linear combinations of atomic
orbitals (LCAO)\cite{Ask_B09}. Within the PAW formalism one works implicitly with
the all-electron wave functions and has access to the (frozen) core
states. This makes the method applicable to a very broad range of systems
including materials with strongly localized $d$ or $f$ electrons which
can be problematic to describe with pseudopotentials. An additional
advantage of the PAW formalism, with respect to linear response
theory, is that the optical transition operator in the long wavelength
limit can be obtained directly due to the use of all-electron
wavefunctions\cite{Bechstedt_B01}. The non-interacting response
function, $\chi^0$, is built from the single-particle eigenstates
obtained either on a real space grid, which is the standard
representation in the GPAW code, or in terms of a localized atomic
orbital (LCAO) basis. We have found that the latter choice reduces the
computational cost of $\chi^0$ considerably while still preserving
the high accuracy of the grid calculation.

The method is used to calculate the macroscopic dielectric constants
of a number of bulk semiconductors, showing very good agreement with
previous calculations as well as experiments. For the surface plasmons
of the Mg(0001) surface we find, in agreement with previous studies,
that the ALDA kernel lowers the plasmon energies by around 0.3 eV
relative to the RPA values and thereby reduces the deviation from
experiments from 4$\%$ to 1-2$\%$. Very good agreement with
experiments is also found for the plasmon energies of graphene which
are shown to exhibit a linear dispersion with a value of 4.9 eV in the
long wave length limit. The deposition of graphene on a SiC substrate
is shown to have little effects on the plasmon energies but leads to
significant broadening of the plasmon resonances. In contrast
deposition on an Al surface completely quenches the graphene plasmons
due to strong non-local electronic screening.

The rest of this paper is organized as follows. Section II introduces the
theoretical framework, where the PAW methodology, the density response  function for
both finite $\ve q$ and $\ve q \rightarrow 0$, and the ALDA kernel in the PAW method
are discussed. The details of the implementation and parallelization in
\textsc{gpaw} and other technical details are presented in section III. Section IV
presents applications for optical properties and plasmon excitations of bulk and
surfaces, where comparison with other calculations and experiments are given. Our
recent investigation on the effect of a semiconducting and metallic substrate on the
plasmon excitations in graphene is also briefly discussed 
 in this section. Finally, a summary is given in section V.  

\section{Method}
\subsection{Basics of the PAW formalism }
In the PAW formalism\cite{Blochl_B94,Blochl_03}, a true all-electron Kohn-Sham
wavefunction $\psi_{n\kk}$ is obtained by a linear transformation
 from a smooth pesudo-wave-function $\tilde \psi_{n\kk}$ via
$\psi_{n\kk}=\hat{\mathcal{T}} \tilde \psi_{n\kk}$. 
 The transformation operator is chosen in such a way that the all-electron
wavefunction $\psi_{n\kk}$ is the sum of the pseudo one $\tilde \psi_{n\kk}$ and an
additive contribution centered around each atom written as
\begin{equation}
\label{Eq:pseudo_vs_AE_wfs}
 \psi_{n\kk}(\rr) = \tilde \psi_{n\kk}(\rr) + \sum_{a,i} 
\langle \tilde p_i^a | \tilde \psi_{n\kk}\rangle 
[\phi_i^a(\rr-\ve R_a) - \tilde \phi_i^a(\rr-\ve R_a)]  
\end{equation}
The pseudo-wave-function $\tilde \psi_{n\kk}$  matches the all-electron one
$\psi_{n\kk}$ outside the augmentation spheres centered on each atom $a$ at position
$\ve R_a$. Their differences inside the augmentation region are expanded on
atom-centered all-electron partial waves $\phi_i^a$ and the smooth counterparts
$\tilde \phi_i^a$. The expansion coefficient is given by $\langle \tilde p_i^a |
\tilde \psi_{n\kk}\rangle $, where $\tilde p_i^a$ is chosen as a dual basis to the
pseudo-partial wave and is called a projector function. A frequently occuring term
is the all-electron expectation value for a semilocal operator $A$ written as
\begin{align}\label{Eq:O}
 & \langle \psi_{n\kk} | A |  \psi_{n\kk} \rangle
 = \langle \tilde \psi_{n\kk} | A |  \tilde \psi_{n\kk} \rangle
\nonumber \\
  &  + \sum_{a,ij} \langle \tilde \psi_{n\kk} | \tilde p_i^a \rangle 
                         \langle \tilde p_j^a | \tilde \psi_{n\kk} \rangle 
                         [\langle \phi_i^a | A | \phi_j^a \rangle 
                           - \langle \tilde\phi_i^a | A | \tilde\phi_j^a \rangle]
\end{align}

\subsection{Density response function and dielectric matrix}
A key concept in TDDFT is the density response function $\chi$. It is defined as
$\chi(\rr, \rp, \w) = \delta n(\rr, \w) / \delta{V_{\mathrm{ext}}(\rp, \w)}$, where
$V_{\mathrm{ext}}$ is the external perturbing potential and  $\delta n$ is the
induced density under the perturbation. For periodic systems, $\chi$ can be written
in the form
\begin{equation}
  \chi(\mathbf{r}, \mathbf{r}^{\prime},  \omega) = \frac{1}{N_q\Omega} 
  \sum_{\mathbf{q}}^{\mathrm{BZ}} \sum_{\mathbf{G} \mathbf{G}^{\prime}}
  e^{i(\mathbf{q} + \mathbf{G}) \cdot \mathbf{r}} \chi_{\mathbf{G}
\mathbf{G}^{\prime}}(\mathbf{q}, \omega) 
  e^{-i(\mathbf{q} + \mathbf{G}^{\prime}) \cdot \mathbf{r}^{\prime}},
\end{equation}
where $\ve G, \ve G^{\prime}$ are reciprocal lattice vectors, $\ve q$ is a wave
vector restricted to the first Broullion Zone (BZ), $N_q$ is the number of $\ve q$
vectors
and $\Omega$ is the volume of the real space primitive cell.

The density response function of the interacting electron system, $\chi$, can be
obtained from the non-interacting density response
function of the Kohn-Sham system, $\chi^0$, and a kernel, $K$, describing the
electron-electron interactions by solving a Dyson-like equation  
\begin{align}
\label{Eq: Dyson_GG}
 & \chi_{\mathbf G \mathbf G^{\prime}}(\mathbf q, \omega)  
  = \chi^0_{\mathbf G \mathbf G^{\prime}}(\mathbf q, \omega) \nonumber \\
 & + \sum_{\mathbf G_1 \mathbf G_2} \chi^0_{\mathbf G \mathbf G_1}(\mathbf q,
\omega) K_{\mathbf G_1 \mathbf G_2}(\mathbf q)
  \chi_{\mathbf G_2 \mathbf G^{\prime}}(\mathbf q, \omega). 
\end{align}

The expression for the non-interacting density response function in the Bloch
representation of Adler and Wiser\cite{Adler_B62,Wiser_B63}, is 
\begin{eqnarray}\label{chi0_GG}
 \chi^0_{\mathbf{G} \mathbf{G}^{\prime}}(\mathbf{q}, \omega) &=& \frac{2}{\Omega} 
 \sum_{\ve k, n n^{\prime}} (f_{n\mathbf{k}}-f_{n^{\prime} \mathbf{k} + \mathbf{q}
}) \nonumber\\
 & \times  &
 \frac{n_{n\ve k,\np \kq}(\ve G) n^{\ast}_{n\ve k,\np \kq}(\ve G^{\prime})}{\omega +
\epsilon_{n\mathbf{k}} - \epsilon_{n^{\prime} \mathbf{k} + \mathbf{q} } + i\eta},
\end{eqnarray}
where
\begin{equation}
\label{Eq:density_matrix}
n_{n\ve k,\np \kq}(\ve G)
\equiv \langle \psi_{n\kk} | e^{-i (\ve q +\ve G)\cdot \rr} | \psi_{n^{\prime} \kq}
\rangle 
\end{equation}
is defined as the charge density matrix. Its evaluation within the PAW formalism is
explained in detail in the
following subsection.
$\epsilon_{n\ve k}$, $f_{n \ve k}$ and $\psi_{n\ve k}$ are the Kohn-Sham eigen-energy, occupation and wave function for band index $n$ and wave vector $\ve k$, and
$\eta$ is a broadening parameter. The summation over $\ve k$ runs all over the BZ
and  $\sum_{\ve k} f_{n\ve k} = 1$ is satisfied for the occupied states. The factor of 
2 accounts for spin (we assume a spin-degenerate system). 

The kernel in Eq. (\ref{Eq: Dyson_GG}) consists of both a Coulomb and an
exchange-correlation(xc) part. The Coulomb kernel is diagonal in the Bloch
representation and written as
\begin{equation}\label{eq.coulomb}
  K^{\mathrm{C}}_{\mathbf G_1 \mathbf G_2}(\mathbf q) = 
  \frac{4\pi}{|\mathbf q+\mathbf G_1|^2} \delta_{\mathbf G_1 \mathbf G_2}, 
\end{equation}
while the xc kernel evaluated within ALDA is given by  
\begin{equation}
\label{Eq:ALDA_kernel}
 K_{\ve G_1 \ve G_2}^{\mathrm{xc-ALDA}} (\ve q) = \frac{1}{\Omega} \int d\ve r
f_{\mathrm{xc}}[n(\ve r)] 
e^{-i(\ve G_1 - \ve G_2) \cdot \ve r},
\end{equation}
with
\begin{equation}
 f_{\mathrm{xc}}[n(\ve r)] = \left. \frac{\partial^2 E_{\mathrm{xc}}[n]}{\partial
n^2} \right |_{n_0(\ve r)}. 
\end{equation}
Details on the evaluation of the xc kernel in the PAW method can be found in a
following subsection.

The Fourier transform of the microscopic dielectric matrix, defined as
$\epsilon^{-1}(\rr, \rp, \w) = \delta V_{\mathrm{tot}}(\rr, \w) /
\delta{V_{\mathrm{ext}}(\rp, \w)}$, is 
related to the density response function via 
\begin{equation}
  \epsilon^{-1}_{\mathbf G \mathbf G^{\prime}}(\mathbf q, \omega)
  = \delta_{\mathbf G \mathbf G^{\prime}} + \frac{4\pi}{|\mathbf q + \mathbf G|^2} 
  \chi_{\mathbf G \mathbf G^{\prime}}(\mathbf q, \omega)
\end{equation}
where $\chi$ is obtained from $\chi^0$ according to Eq. (\ref{Eq: Dyson_GG}).
The off-diagonal elements of the $\chi^0_{\ve G \ve G^{\prime}}$ matrix describes the
response of the electrons at wave vectors different from the external perturbing
field and thus contain information about the inhomogeneity of the microscopic
response of electrons known as the 'local field effect'\cite{Louie_B87}. The
macroscopic dielectric function is defined as
\begin{equation}
 \epsilon_M(\ve q, \omega) = \frac{1}{\epsilon^{-1}_{\ve 0 \ve 0}(\ve q,
\omega)},
\end{equation}
and is directly related to many experimental
properties. For example, the optical absorption spectrum (ABS) is given by
$\mathrm{Im} \epsilon_M(\ve q \rightarrow 0, \omega)$. The electron energy loss spectrum (EELS\cite{Ibach}) is propotional to
$-\mathrm{Im} (1/\epsilon_M)$. Both spectra reveal information about the elementary
electronic excitations of the system. EELS is especially useful in probing
the collective electronic excitations, known as plasmons,  of bulk and
low-dimensional systems\cite{Ibach}.

\subsection{Charge density matrix in the PAW method}
In this subsection, we will discuss the charge density matrix 
$n_{n\ve k,\np \kq}(\ve G)$, which is defined in Eq. (\ref{Eq:density_matrix}) and
is a  crucial quantity for the evaluation of $\chi^0$. Care must be taken for the
long wavelength limit ($\ve q \rightarrow 0$) since the Coulomb kernel, $4\pi / |\ve
q + \ve G|^2$, diverges at $\ve q \rightarrow 0$ and $\ve G = 0$; while the charge
density matrix approaches zero at this limit. As a result, we separate the
discussion into two parts: finite $\ve q$ and $\ve q \rightarrow 0$.

\subsubsection{Finite q}
Considering the transformation between the pseudo-wavefunction and the all-electron
wavefunction in Eq. (\ref{Eq:pseudo_vs_AE_wfs}) and employing  Eq. (\ref{Eq:O})
yields
\begin{eqnarray}
\label{Eq:density_matrix_finteq}
   n_{n\ve k,\np \kq}(\ve G) &=&  \tilde{n}_{n\ve k,\np \kq}(\ve G)\\ 
 & + & \sum_{a,ij} 
   \langle \tilde{\psi}_{n \mathbf k}  |  \tilde{p}_i^a \rangle
   \langle  \tilde{p}_j^a | \tilde{\psi}_{n^{\prime} \mathbf k + \mathbf q}  \rangle
Q^a_{ij}(\ve q + \ve G) \nonumber
\end{eqnarray}
with
\begin{eqnarray}
\label{Eq:pseudo_density_matrix_finiteq}
\tilde{n}_{n\ve k,\np \kq}(\ve G) \equiv 
\langle \tilde{\psi}_{n \mathbf k} | 
   e^{-i (\mathbf q + \mathbf G) \cdot \mathbf r} | \tilde{\psi}_{n^{\prime} \mathbf
k + \mathbf q} \rangle \\
\label{Eq:PAW_density_matrix_finiteq}
 Q^a_{ij}(\ve K) \equiv \langle \phi_i^a | e^{-i\mathbf{K} \cdot \mathbf{r}} |
\phi_j^a \rangle
         - \langle \tilde{\phi}_i^a | e^{-i \mathbf{K} \cdot \mathbf{r}} |
\tilde{\phi}_j^a \rangle
\end{eqnarray}
and $\ve K \equiv \ve q + \ve G$. 

The pseudo-density matrix in Eq. (\ref{Eq:pseudo_density_matrix_finiteq}) is
calculated using a mixed space scheme. First, the cell periodic function $\tilde
\psi^{\ast}_{n\kk}(\rr) 
\tilde \psi_{n^{\prime}\ve k+\ve q}(\rr) e^{-i\ve q \cdot \rr}$ is evaluated on
a real-space grid; then it is Fourier transformed to get 
\begin{equation}
 \tilde{n}_{n\ve k,\np \kq}(\ve G) 
= \mathcal{F}\left[\tilde \psi^{\ast}_{n\kk}(\rr) 
\tilde \psi_{n^{\prime}\ve k+\ve q}(\rr) e^{-i\ve q \cdot \rr}\right]
\end{equation}

The augmentation part in Eq. (\ref{Eq:PAW_density_matrix_finiteq}) is calculated on
fine one-dimensional radial grids centered on each atom.
Such fine grids are required to represent accurately the oscillating nature of the
all-electron partial wave in 
the augmentation region. 
The plane wave term $e^{-i\ve K \cdot \rr}$ is expanded using real spherical
harmonics by 
\begin{equation}
    e^{-i \mathbf{K} \cdot \mathbf{r}} = 4 \pi \sum_{lm} (-i)^l j_l(|\ve K|r)
Y_{lm}(\hat{\mathbf{r}})  Y_{lm}(\hat{\mathbf{K}}), 
\end{equation}
where $j_l$ is spherical Bessel function for angular momentum $l$ and
$\hat{\mathbf{K}} = \mathbf{K} / |\ve K|$. 
Combining the above equations and the expression for the partial wave $|\phi_i^a
\rangle= \phi_{n_il_i}^a(r) Y_{l_im_i}(\hat{\rr})$, we can write
\begin{align}
    & Q_{ij}^a(\ve K)
    = 4 \pi e^{-i \mathbf{K} \cdot \mathbf{R}_a}  \sum_{lm} (-i)^l 
Y_{lm}(\hat{\mathbf{K}}) 
     \int d\hat{\rr} \  Y_{lm} Y_{l_i m_i} Y_{l_j m_j}  \nonumber     \\
    & \times 
        \int dr \ r^2  j_l(|\ve K|r) \left[ \phi^{a}_{n_i l_i}(r)  \phi^{a}_{n_j
l_j}(r) 
                                     -  \tilde{\phi}^{a}_{n_i l_i}(r) 
\tilde{\phi}^{a}_{n_j l_j}(r) \right]  
\end{align}

\subsubsection{Long wave length limit}
In the long wave length limit, the $\ve G \neq 0$ components of the density matrix
$n_{n\ve k,\np \kq}(\ve G)$ remain the same as that for finite $\ve q$. Only the
$\ve G = 0$ components need to be modified and are written as
\begin{equation}\label{Eq:dipole_optical}
 n_{n\ve k,\np \ve k + \ve q}( 0) |_{\ve q \rightarrow 0} \equiv \langle \psi_{n\kk}
| e^{-i \ve q \cdot \rr} | \psi_{n^{\prime} \kq} \rangle_{\ve q \rightarrow 0}.
\end{equation}
In Ref. \onlinecite{Kresse_B06}, the above so called longitudinal form is derived in
the PAW framework by using Taylor expansion of the $e^{i\ve q \cdot \ve r}$ to the
first order. 
Here we adopt an alternative but equivalent form which can be
derived using the second order $k \cdot p$ perturbation theory\cite{Guiseppe} as
described below. 

Expressing the wavefunction using Bloch's theorem as $\psi_{n\ve k}(\ve r)=u_{n \ve
k}(\ve r) e^{i\ve k \cdot \ve r}$, where $u_{n\ve k}(\ve r)$ is the periodic Bloch
wave, the dipole transition element in Eq. (\ref{Eq:dipole_optical}) becomes 
\begin{equation}\label{Eq:Bloch_transition_element}
 \langle \psi_{n\kk} | e^{-i \ve q \cdot \rr} | \psi_{n^{\prime} \kq} \rangle
 = \langle u_{n\kk} |  u_{n^{\prime} \kq} \rangle.
\end{equation}
For vanishing  $\ve q$, the wavefunction for $ | u_{n^{\prime} \kq} \rangle$ can be
obtained in terms of those for $ | u_{m \kk} \rangle$ through second order
perturbation theory:
\begin{equation}
  | u_{n^{\prime} \mathbf k + \mathbf q} \rangle
   =  | u_{n^{\prime} \mathbf k } \rangle
   +   \sum_{m \neq n^{\prime}} 
   \frac{ \langle \psi_{m \mathbf k} | \tilde V | u_{n^{\prime} \mathbf k} \rangle
}{\epsilon_{n^{\prime} \mathbf k} - \epsilon_{m \mathbf k} } | u_{m \mathbf k}
\rangle
\end{equation}
The perturbing potential $\tilde V$ in the above equation is obtained through
\begin{equation}\label{Eq:perturbing_V}
 \tilde V = H(\ve k + \ve q) - H(\ve k) = -i \ve q \cdot (\ve{\nabla} + i \ve k),
\end{equation}
where 
\begin{equation}
 H(\ve k) = -\frac{1}{2}(\ve{\nabla} + i\ve k)^2 + V(\ve r)
\end{equation}
is the $k \cdot p$ hamiltonian\cite{Guiseppe} and $V(\ve r)$ is the effective
Kohn-Sham potential.

Combining Eq. (\ref{Eq:Bloch_transition_element}) - (\ref{Eq:perturbing_V}),  the
charge density matrix at the long wavelength limit becomes
\begin{eqnarray}\label{Eq:density_matrix_q0}
 n_{n\ve k,\np \ve k + \ve q}( 0) |_{\ve q \rightarrow 0} 
&=&
\frac{-i\ve q \cdot \langle n_{n \mathbf k} |  \ve{\ve{\nabla}} + i\ve k |
u_{n^{\prime} \mathbf k} \rangle }{\epsilon_{n^{\prime} \mathbf k} - \epsilon_{n
\mathbf k} }, \nonumber\\
&=& \frac{-i\ve q \cdot \langle \psi_{n \mathbf k} |  \ve{\ve{\nabla}} |
\psi_{n^{\prime} \mathbf k} \rangle }{\epsilon_{n^{\prime} \mathbf k} - \epsilon_{n
\mathbf k} }.
\end{eqnarray}
The above expression for the charge density matrix in the PAW method has
an advantage over the pseudopotential method, where the nabla operator has to be
corrected by the
commutator of the non-local part of pseudopotential with the position operator $\rr$
\cite{Bechstedt_B01}. 
In the PAW method, the matrix element $\langle \psi_{n \mathbf{k}} | \ve{\nabla} |
\psi_{n^{\prime} \mathbf{k}} \rangle$ is given by 
\begin{align}
     & \langle \psi_{n \mathbf{k}} | \ve{\nabla} | \psi_{n^{\prime} \mathbf{k}} \rangle
     = \langle \tilde{\psi}_{n \mathbf{k}} | \ve{\nabla} | \tilde{\psi}_{n^{\prime}
\mathbf{k}} \rangle
\nonumber \\     
  &+  \sum_{a,ij} 
   \langle  \tilde{\psi}_{n \mathbf k} | \tilde{p}_i^a  \rangle
   \langle \tilde{p}_j^a | \tilde{\psi}_{n^{\prime} \mathbf k}   \rangle
   \left[ \langle \phi_i^a | \ve{\nabla} | \phi_j^a \rangle
         - \langle \tilde{\phi}_i^a | \ve{\nabla} | \tilde{\phi}_j^a \rangle
   \right],
\end{align}
In GPAW, where the pseudo wave functions, $\tilde \psi_{n\kk}$, are represented on a
real space grid, the first matrix element is calculated using a finite difference
approximation for the
nabla operator. The augmentation part is evaluated on fine one dimensional radial
grids. The
nabla operator combined with partial waves $\phi_i^a(\rr) = \phi^a_{n_1 l_1}(r)
Y_{l_1 m_1}(\hat{\rr})$ and $\phi_j^a(\rr) = \phi^a_{n_2 l_2}(r) Y_{l_2
m_2}(\hat{\rr})$ is written as 
\begin{align}\label{Eq:nabla}
  & \langle \phi_i^a | \ve{\nabla} | \phi_j^a \rangle 
\nonumber \\ 
&= \langle \phi_i^a | \displaystyle \frac{\partial}{\partial r}(\frac{\phi^a_{n_2
l_2}}{r^{l_2}} ) \frac{\partial r}{\partial \rr}
        r^{l_2}Y_{l_2 m_2} \rangle 
    + \langle \phi_i^a | \frac{\phi^a_{n_2 l_2}}{r^{l_2}} \ve{\nabla} (r^{l_2}Y_{l_2
m_2}) \rangle. 
\end{align}
Since real spherical harmonics are employed, we get 
\begin{equation}
 \frac{\partial r}{\partial \rr} = (\frac{x}{r}, \frac{y}{r},\frac{z}{r}) 
  = \sqrt{\frac{4\pi}{3}} (Y_{1m_x}, Y_{1m_y}, Y_{1m_z})
\end{equation}
Substitute the above equation into Eq. (\ref{Eq:nabla}) and split the integration
into radial and angular parts,we get for the x-component
\begin{align}
& \langle \phi_i^a | \frac{\partial}{\partial x} | \phi_j^a \rangle
\nonumber \\
 &= \sqrt{\frac{4\pi}{3}} \int dr \ r^2 \phi^a_{n_1 l_1} 
     \frac{\partial}{\partial r}(\frac{\phi^a_{n_2 l_2}}{r^{l_2}} ) r^{l_2} 
      \int d\hat{\rr} \ Y_{l_1 m_1} Y_{l_2 m_2} Y_{1 m_x} \nonumber \\
 &+ \int dr \ r^2  \phi^a_{n_1 l_1} \frac{\phi^a_{n_2 l_2}}{r}
     \int d\hat{\rr} \ Y_{l_1 m_1} r^{1-l_2} \frac{\partial}{\partial x}
(r^{l_2}Y_{l_2 m_2})
\end{align}
The derivation for the y- and z-component and for the pseudo-partial-wave follows in
a similar way.  

\subsection{The ALDA xc kernel in the PAW method}
The ALDA xc kernel, expressed in Eq. (\ref{Eq:ALDA_kernel}), is evaluated using the
all-electron density, which takes the form 
\begin{equation}
 n(\ve r)  = \tilde{n}(\ve r) + \sum_a [n^a(\ve r - \ve R_a) - \tilde{n}^a(\ve r -
\ve R_a)],
\end{equation}
where
\begin{eqnarray}
\label{Eq:Pseudo_density}
 \tilde{n}(\ve r) &=& \sum_{n \ve k} f_{n \ve k} |\tilde{\psi}_{n\ve k} (\ve r)|^2 
+ \sum_a \tilde{n}_c^a(|\ve r - \ve R_a|), \\
 n^a(\ve r) &=& \sum_{ij} D_{ij}^a \phi_i^a(\ve r) \phi_j^a(\ve r) + n^a_c(\ve r), \\
 \tilde{n}^a(\ve r) &=& \sum_{ij} D_{ij}^a \tilde{\phi}_i^a(\ve r)
\tilde{\phi}_j^a(\ve r)  
  + \tilde{n}^a_c(\ve r),
\end{eqnarray}
with $D_{ij}^a = \sum_{n\ve k} \langle  \tilde{\psi}_{n \mathbf k} | \tilde{p}_i^a 
\rangle
f_{n\ve k} \langle \tilde{p}_j^a | \tilde{\psi}_{n \mathbf k}   \rangle$. Here
$n_c^a(\ve r)$ is the 
all-electron core density and $\tilde{n}_c^a(\ve r)$ can be chosen as any smooth
continuation of 
$n_c^a(\ve r)$ inside the augmentation sphere since it will be canceled out in Eq.
(\ref{Eq:Pseudo_density}).

The ALDA xc kernel can also be separated into smooth and atom-centered contributions
\begin{equation}
 K_{\ve G_1 \ve G_2}^{\mathrm{xc-ALDA}} = \tilde{K}_{\ve G_1 \ve
G_2}^{\mathrm{xc-ALDA}}
 + \sum_a \Delta K_{\ve G_1 \ve G_2}^{a, \mathrm{xc-ALDA}}.
\end{equation}
The smooth part is constructed from pseudo-density and by utilizing a Fourier
transform 
\begin{eqnarray}
 \tilde{K}_{\ve G_1 \ve G_2}^{\mathrm{xc-ALDA}}
&=& \frac{1}{\Omega} \int d \ve r f_{\mathrm{xc}}[\tilde{n}(\ve r)] 
e^{-i(\ve G_1 - \ve G_2) \cdot \ve r} \nonumber \\
&=& \frac{1}{\Omega} \mathcal{F} \left.\left\{ f_{\mathrm{xc}}[\tilde{n}(\ve r)] 
\right\} 
\right|_{\ve G_1 - \ve G_2} 
\end{eqnarray}
The atom-centered contribution is evaluated on 1D grids
\begin{eqnarray}
 \Delta K_{\ve G_1 \ve G_2}^{a, \mathrm{xc-ALDA}} &=& 
\frac{1}{\Omega}  \int r^2 dr d \hat{\ve r}
e^{-i(\ve G_1 - \ve G_2) \cdot \ve r} \nonumber \\
& & \times
[ f_{\mathrm{xc}}[n^a] - 
        f_{\mathrm{xc}}[\tilde{n}^a]] 
\end{eqnarray}

\section{Numerical details}
In this section we describe the most important numerical and technical
aspects of our implementation; in particular the Hilbert transform
used to obtain $\chi^0$ from the dynamic form factor (spectral
function) and the applied parallelization scheme.

\subsection{Symmetry}
For each wave vector $\ve q$, the evaluation of $\chi^0$ involves a summation over
occupied and empty states in the entire
BZ. By exploiting the crystal symmetries, however, we need only calculate the wave
functions and energies in the irreducible BZ. 
This is because the wave function at a general $k$-point can always be obtained from
a wave function in the irreducible part of BZ by application of a symmetry
transformation, $T$. In general we have the relation 
\begin{equation}
\psi_{n,T \ve k}(\ve r)=\psi_{n,\ve k}(T^{-1}\ve r)
\end{equation}
where $\ve k$ belongs to the IBZ. The above relation can be directly
verified by considering how the right hand side transforms under
lattice translations. In addition to the crystal symmetries, time reversal symmetry
applies to any system in the absence of magnetic fields
\begin{equation}
\psi_{-\ve k}(\ve r) = \psi^{\ast}_{\ve k}(\ve r)
\end{equation}

\subsection{Hilbert transform}
Rather than constructing $\chi^0$ directly from Eq. (\ref{chi0_GG}) we obtain it as
a Hilbert transform of the (non-interacting) dynamic form factor,
$S^0$.\cite{Yambo,Dresponse} The latter is given by 
\begin{eqnarray}\label{Eq:S_GG}
S^0_{\mathbf{G} \mathbf{G}^{\prime}}(\mathbf{q}, \omega) &=& \frac{2}{\Omega} 
 \sum_{\ve k, n n^{\prime}} (f_{n\mathbf{k}}-f_{n^{\prime} \mathbf{k} + \mathbf{q} })
 \delta(\omega + \epsilon_{n\ve k} - \epsilon_{n^{\prime}\ve k+\ve q})
 \nonumber\\
& \times & 
 n_{n\ve k,\np \kq}(\ve G) n^{\ast}_{n\ve k,\np \kq}(\ve G^{\prime}).
\end{eqnarray}
In practice $S^0(\omega)$ is evaluated on a uniform frequency grid
extending from 0 to around 40-60 eV with a grid spacing in the range 0.01-0.1~eV,
and the delta functions are approximated by triangular functions following Ref.
\onlinecite{Kresse_B06_GW}. The non-interacting response function is obtained as
\begin{eqnarray}\label{Eq:Hilbert}
 \chi^0_{\ve G \ve G^{\prime}}(\ve q, \omega) &=& \int_{0}^{\infty} d \omega^{\prime}
S^0_{\ve G \ve G^{\prime}}(\ve q, \omega^{\prime})  \nonumber \\
&\times&
\left[ \frac{1}{\omega - \omega^{\prime}+i\eta}
-\frac{1}{\omega+\omega^{\prime}+i\eta}  \right].
\end{eqnarray}
The above Hilbert transform is performed directly on the frequency
grid setting the broadening parameter $\eta$ equal to the grid
spacing.

\subsection{LCAO vs grid calculations}
It is well known that the use of localized atomic orbitals as basis
functions can significantly reduce the computational effort of
groundstate electronic structure calculations. For calculations of the
density response function the use of localized basis functions is
complicated by the fact such basis sets are typically not closed under
multiplication\cite{Sham_B75,Gunnarsson_B94,Schattke_B02}. As a
consequence the size of the product basis needed to represent the
response function grows as $N_{\mu}^2$, where $N_\mu$ is the number of
basis functions used to represent the wave functions (we note that for strictly
localized basis functions, the effective size of the ``product basis''
grows only linearly with the system size because pair densities of
non-overlapping orbitals vanishes, however, the prefactor is typically very large). 
A further challenge is the computation of the Coulomb interaction kernel,
$1/|\ve r - \ve \rp|$, in the product basis leading to six-dimensional multi center
integrals. These intergrals must be performed
either by using efficient Poisson solvers or by resorting to analytical
techniques. The latter is extensively used in quantum chemistry codes
applying Gaussian basis sets.

\begin{figure}[t]
    \centering
    \includegraphics[width=1.0\linewidth,angle=0]{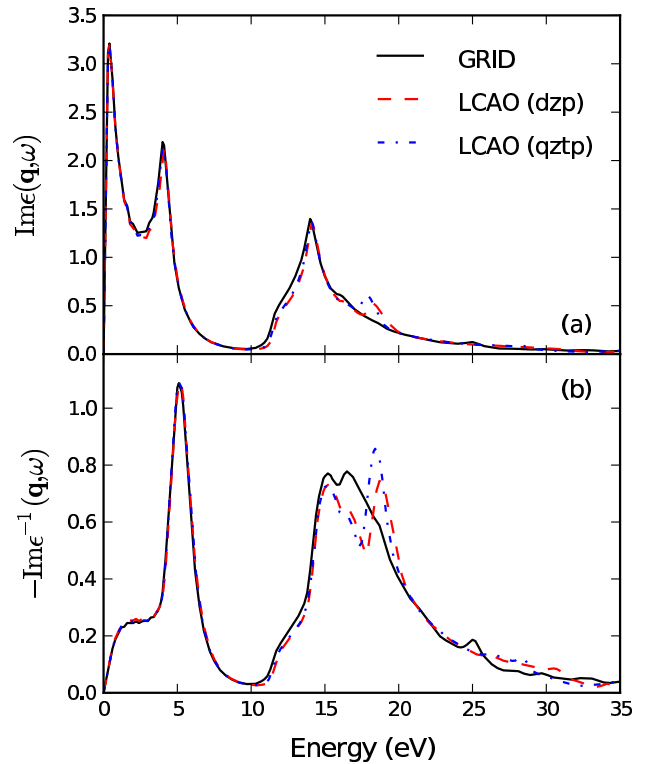}
    \caption{(Color online) The imaginary part of the dielectric function (a) and
energy loss
function (b) of graphene at $q = 0.046$ \AA$^{-1}$ along $\bar{\Gamma}-\bar{M}$
direction of its surface Broullion zone (SBZ) calculated with 3D uniform grid
(GRID, black solid line) and localized atomic orbital (LCAO) using dzp (red
dashed line) and qztp (blue dash-dotted line) basis, respectively . }
    \label{fig:grid_vs_lcao_graphene}
\end{figure}

For these reasons we have chosen to represent the density response
function in a plane wave basis. The plane wave basis is closed under
multiplication and the Coulomb kernel is simply given by Eq.
(\ref{eq.coulomb}). However, we still keep the advantage of using an
LCAO as basis in the calculation of the Kohn-Sham wave functions and
energies which enter the construction of $\chi^0$.\cite{Ask_B09} Apart
from reducing the computational effort of the groundstate calculation
(which must include many unoccupied bands), the storage requirements
for wave functions become much less than for corresponding grid or
plane wave calculations. This is because the LCAO coefficients provide
a more compact representation of the wave functions, in particular for open structures
containing large vacuum regions, and because significantly fewer unoccupied wave
functions result from the LCAO calculation (for a fixed energy cut-off).

Compared to plane waves or real space grids, LCAO calculations
employing standard basis sets usually give a less accurate but often acceptable 
description of the occupied and low-lying unoccupied wave functions
and energies. For higher-lying unoccupied states, blue shifts are
expected due to the (unphysical)confinement imposed by the localized
basis set, and the continuum is broken into discrete bands. Despite these effects,
we have found that the use of LCAO wave functions instead of grid wave functions
has rather little effect on the dielectric function -- at least in the relevant low
energy regime.

As an example Fig. \ref{fig:grid_vs_lcao_graphene} shows the absorption spectrum (a)
and EELS spectrum (b) of graphene calculated using wave functions and energies from
a grid calculation and from an LCAO with double-zeta polarized (dzp) and quadruple
zeta triple polarized
(qztp) basis, respectively. The unit cell is the primitive cell of graphene
containing two carbon atoms and with 20
\AA~vacuum. The BZ is sampled on a $64\times 64$  Monkhorst-Pack grid. The number of
bands included are 60 for the grid and LCAO(qztp) basis and 26 for the
LCAO(dzp) basis. In all three cases this corresponds to inclusion of states with
energy below 40 eV.
The response function is evaluated at the RPA level including local field effects up
to a plane wave cut-off of 150 eV. 

\begin{figure}[t]
    \centering
    \includegraphics[width=1.0\linewidth,angle=0]{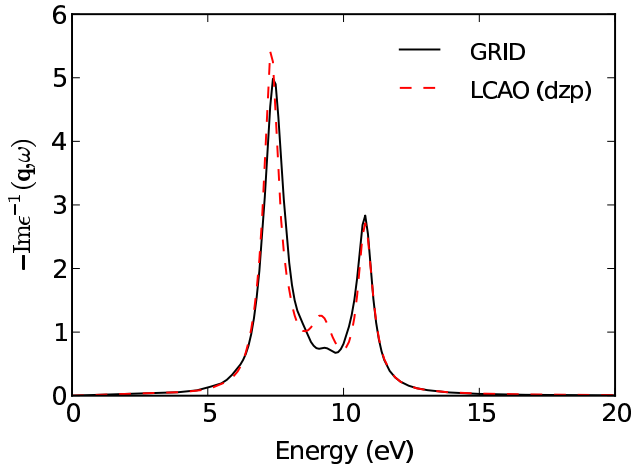}
    \caption{(Color online) The energy loss function  of
Mg(0001) surface  at $q =0.07$ \AA$^{-1}$ along $\bar{\Gamma}-\bar{M}$ direction of
its 
SBZ calculated with 3D uniform grid (GRID, black solid line) and
localized atomic orbital (LCAO) using dzp (red dashed line) basis. 
The dzp basis used here includes double-zeta orbitals of 3s and 3p atomic orbitals
as well as
one d-type Gaussian polarization function.}
    \label{fig:grid_vs_lcao_Mg}
\end{figure}

For excitation energies below 10 eV, the LCAO results agree remarkably
well with the grid calculations. The $\pi \rightarrow \pi^{\ast}$
absorption peak at around 4 eV in panel (a) and the $\pi$ plasmon
around 5 eV in panel (b) are well reproduced in LCAO calculations. For
energies above 10 eV, we observe slight deviations, however, the
overall agreement is remarkable for the entire energy range.
In particular, the $\sigma \rightarrow
\sigma^{\ast}$ transition at around 14 eV in panel (a) and the $\sigma$ plasmon
around 17 eV in panel (b) are clearly visible, although in the LCAO calculation the
latter is splitted into two peaks.

Fig. \ref{fig:grid_vs_lcao_Mg} shows another example of the Mg(0001)
surface, which is modeled by a slab of 16 layer. The energy loss
function calculated with 3D grids is characterized by 
two peaks at around 7.5 and 11 eV, which correspond to the surface and
bulk plasmons, respectively.   Again the LCAO(dzp) calculation
reproduces the grid results quite accurately, the only discrepancy
being the slight discrepancy (0.1 eV) of the peak just below 10 eV. 
Note that dzp basis used in this case includes double-zeta orbitals of 3s and 3p
atomic orbitals as well as
one d-type Gaussian polarization function. The inclusion of the d-type orbital in
the basis function is 
crucial for the correct description of both the electronic structure (fx, band
structure and density of states) 
and the surface plasmon of the Mg surface.
The response function calculation presented here is performed at the RPA level, with
summation over bands up to 15 eV and inclusion of local fields up to 50~eV plane wave
cutoff. The frequency grid spacing is 0.1 eV. 
The response function is not fully converged with these parameters.
 Please refer to the next section for the converged
  results for Mg(0001) surface.

\subsection{Storage of wave functions}
For ground state calculations performed using grid based wave
functions, the entire set of occupied and unoccupied wave functions
might be too large to be stored on disk, making the separation of the
ground state and response function calculations impossible.  In this
case, the response function, or more precisely, the dynamical form
factor of Eq. (\ref{Eq:S_GG}), is constructed as the wave functions
are calculated.

In the LCAO mode, only the expansion coefficients of the wave
functions in terms of the localized basis functions are calculated and
stored. Since this representation is significantly more compact than
the grid representation, the entire set of wave functions can be
calculated and stored at once, and the calculation of the response
function can be performed as a post-processing step.

\subsection{Parallelization}
\begin{figure}[t]
    \centering
\includegraphics[width=1.0\linewidth]{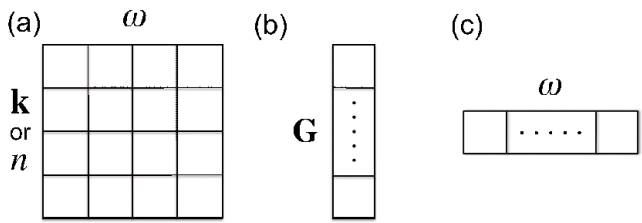}
    \caption{Schematic illustration of the applied parallelization
      scheme. Each box represents a single CPU. (a) The calculation of
      $S^0_{\ve G \ve G^{\prime}}(\ve q, \omega)$ is performed in
      parallel over wave vectors, $\ve k$, (or bands, $n$, for large
      cells) and frequencies $\omega$. (b) The Hilbert transform is
      parallelized over $\ve G$. (c) Finally the Dyson equation is
      solved by parallelizing over the frequencies.}
    \label{fig:parallelization_figure}
\end{figure}

The calculation of the response function involves the three steps:
evaluation of the spectral function $S^0_{\ve G \ve G^{\prime}}(\ve q,
\omega)$ according to Eq. (\ref{Eq:S_GG}), Hilbert transform following
Eq. (\ref{Eq:Hilbert}) and solving Dyson's equation Eq.  (\ref{Eq:
  Dyson_GG}). Fig. \ref{fig:parallelization_figure} illustrates the
parallelization scheme applied for each of these three steps.

It is natural to parallelize the evaluation of $S^0_{\ve G \ve
  G^{\prime}}(\ve q, \omega^{\prime})$ over k-points (or bands for few
k-point calculations). On the other hand, the size of the matrix is
often too large to be handled on a single CPU. In such cases each CPU
only calculates $S^0$ on a part of the frequency grid. This leads to
the two-dimensional parallelization scheme illustrated in Fig.
\ref{fig:parallelization_figure}(a). Finally the full $S^0_{\ve G \ve
  G^{\prime}}(\ve q, \omega^{\prime})$ is obtained by summing over
$k$-points, i.e. summing up the columns in Fig.
\ref{fig:parallelization_figure}(a). Since the Hilbert transform
involves a frequency convolution it is convinient to redistribute the
data from parallelization over $\omega$ to over $\ve G$. Finally, the
Dyson equation is done separately for each frequency point and is
therefore parallelized over $\omega$, as shown in panel (c).

\section{Results}
In this section, the density response function method is applied to study the optical
properties and plasmon excitations of solids. They are usually measured by optical
and electron energy loss spectroscopy (EELS), which are related to  $\mathrm{Im}
\epsilon_M$ and  $-\mathrm{Im} [1/ \epsilon_M]$, respectively.  
For extended systems, the two kinds of spectroscopy give quite distinct spectra.
The optical absorption spectrum (ABS) is determined by single-particle excitations while
EELS is dominated by  collective electronic excitations, plasmons, which are defined
as $\epsilon_M \rightarrow 0$. 

\subsection{Optical properties}

\begin{table}[t]
\caption{The static macroscopic dielectric constants $\epsilon$ calculated using the
PAW method on the RPA level without local field (NLF) and including local field (LF)
effect. These values are compared with other PAW calculations \cite{Kresse_B06} and
experiments \cite{Dielectric_constant}.}

\begin{tabular*}{0.95\linewidth}{@{\extracolsep{\fill}} c c c c c c}
  \hline \hline 
  Crystal & $\epsilon^{\mathrm{RPA}}_{\mathrm{NLF}}$ 
          & $\epsilon^{\mathrm{RPA}}_{\mathrm{LF}}$ 
          & $\epsilon^{\mathrm{RPA}}_{\mathrm{NLF}}$ [\onlinecite{Kresse_B06}] 
          & $\epsilon^{\mathrm{RPA}}_{\mathrm{LF}}$ [\onlinecite{Kresse_B06}] 
          & Expt. [\onlinecite{Dielectric_constant}]\\
  \hline
  C & 5.98 & 5.58 & 5.98 & 5.55 & 5.70\\
  Si & 13.99 & 12.58 & 14.04 & 12.68 & 11.90 \\ 
  SiC & 7.18 & 6.58 & 7.29 & 6.66 & 6.52\\
  AlP & 9.04 & 7.83 & 9.10 & 7.88 & 7.54 \\
  GaAs & 15.12 & 13.67 & 14.75 & 13.28 & 11.10\\
  \hline \hline 
\end{tabular*}
\end{table}

Table I shows the calculated RPA static dielectric function in the optical limit for
five semiconductors (C, Si, SiC, AlP, GaAs). We use the same lattice constants as
in Ref.~\onlinecite{Kresse_B06} and a grid spacing of 0.2 \AA.  A Monkhorst-Pack
grid of $12\times 12\times 12$  and 60 unoccupied bands are used. We use a Fermi temperature of 
0.001 eV in the ground state LDA calculation and a broadening parameter ($\eta$) of 0.0001 eV in $\chi^0$. Note that in this case we calculate the static response function directly from Eq. (\ref{chi0_GG}), i.e. we do not use the Hilbert transform. For calculations including 
local field effects, a cutoff of 150 eV is used. The dielectric constants obtained
both with and without local fields agree to within 0.1 with previous PAW
calculations\cite{Kresse_B06}. The only exception is GaAs for which our dielectric constant is 0.4 larger. This deviation could come from differences in the PAW setups for Ga or As.  The inclusion of local fields lowers the dielectric constant by 10-15\% in
agreement with earlier reports\cite{Louie_B86}. These obtained values are, however, 
generally larger than the experimental values due to the underestimated band gaps by LDA. Inclusion of
the ALDA kernel only increases the dielectric constant further, and are not reported here. 

\begin{figure}[t]
    \centering
    \includegraphics[width=1.0\linewidth,angle=0]{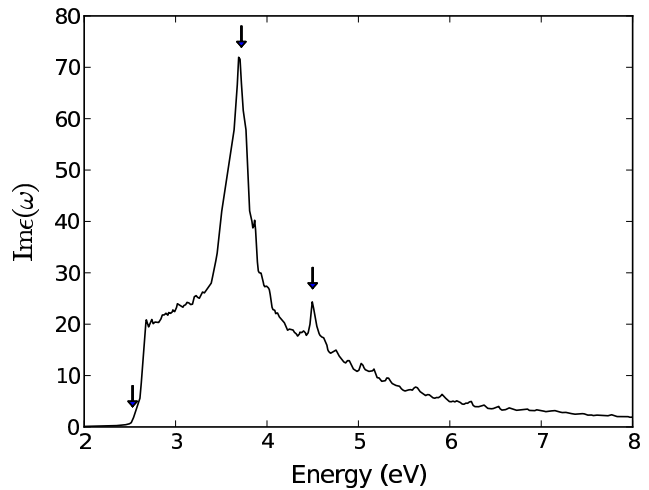}
    \caption{Imaginary part of the dynamical dielectric function of bulk silicon.
The arrows indicate the absorption onset and the position of main and secondary peaks, respectively, as extracted from Ref. \onlinecite{Kresse_B06}.}
    \label{fig:Si_absorption}
\end{figure}

Fig.~\ref{fig:Si_absorption} shows the dynamical dielectric function
for Si.  Compared to the calculations for the static dielectric
constant, a significantly denser k-point sampling of $80 \times 80
\times 80$ is employed here to resolve the finer details in the
spectrum.  A total of 36 unoccupied bands are use in the construction
of $\chi^0$. Local field effects are not included and $\eta$ is set to
0.01 eV. The onset of absorption and the position of the two
characteristic peaks in the absorption spectrum compare very well with
previous RPA calculations\cite{Kresse_B06} as shown by the arrows in the
figure. However, it is quite different from the experimental
absorption spectrum\cite{Aspnes_B83} which exhibits an
absorption onset at $\sim$ 0.5 eV larger than predicted by our
calculation, and shows a double peak around 3.8 eV. The disagreement with the experimental spectrum is
due to the underestimation of the band gap by LDA and the fact that
RPA does not include the electron-hole interaction.

\subsection{Plasmon excitations}
In contrast to the optical excitations, like the Si absorption spectrum discussed in the previous section, plasmon
excitations are generally well described by RPA and TDLDA. 
In the first part of this subsection, a classical calculation of the surface
plasmons of a Mg(0001) surface\cite{Chulkov_L04} is reproduced and the results are
in good agreement with previous reports\cite{Chulkov_L04}. In the second part,  our
recent investigation of the effect of substrates on the plasmonic excitations of
graphene is summarized.

Plasmon excitations appear as strong peaks in the electron energy loss spectrum (EELS) which is directly related to the imaginary inverse dielectric function,
\begin{equation}
 - \mathrm{Im} \epsilon^{-1}(\mathbf{q}, \omega) = -\frac{4\pi}{|\mathbf{q}|^2}
\mathrm{Im} 
 \chi_{\mathbf{G}=0, \mathbf{G}^{\prime}=0}(\mathbf{q}, \omega)
\end{equation}
For excitations at surfaces, a surface loss function can be defined as\cite{Chulkov_L04}
\begin{equation}
 g(\mathbf{q}, \omega) =  
-\frac{2\pi}{|\mathbf{q}|} \iint dz dz^{\prime} 
\chi_{\mathbf{G}_{\parallel}=
\mathbf{G}^{\prime}_{\parallel}=0}(z,z^{\prime};\mathbf{q},\omega)
e^{|\mathbf{q}|(z+z^{\prime})} 
\end{equation}
where $\parallel$ and $z$ correspond to directions parallel and perpendicular to the
surface, respectively, and $\chi_{\mathbf{G}_{\parallel}
\mathbf{G}^{\prime}_{\parallel}}(z,z^{\prime};\mathbf{q},\omega)$
is the Fourier transform of $\chi_{\mathbf{G} \mathbf{G}^{\prime}}(\mathbf{q},
\omega)$ in the $z$-direction. 

\begin{figure}[t]
    \centering
    \includegraphics[width=1.0\linewidth,angle=0]{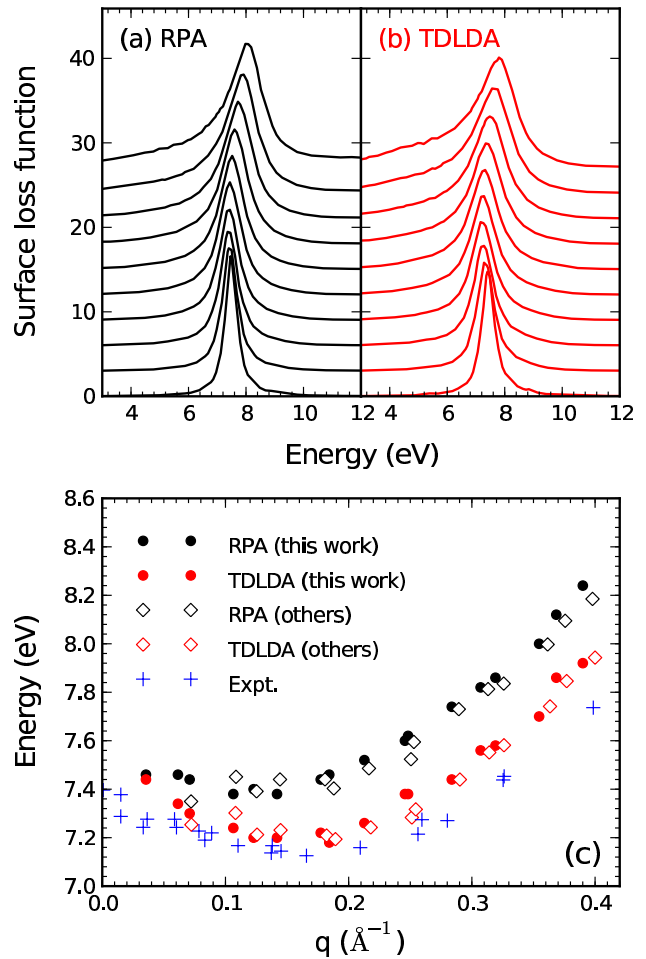}
    \caption{(Color online) Surface loss function of the Mg(0001) surface along
the $\bar{\Gamma} -
\bar{M}$ direction of the surface BZ calculated using RPA (a) and TDLDA (b). In both cases $|\ve q|$ increases 
from bottom to top. (c) Surface plasmon dispersion for both the $\bar{\Gamma} - \bar{M}$ and $\bar{\Gamma} - \bar{K}$ directions. Results from this work (filled dots) compare well with other calculations (hollow dots
\cite{Chulkov_L04}) and experiments\cite{Plummer_SS92}. }
    \label{fig:Mg_fig}
\end{figure}

\subsection{Surface plasmons of Mg(0001)}
Fig. \ref{fig:Mg_fig} shows the surface loss function of the Mg(0001)
surface along the $\bar{\Gamma} - \bar{M}$ direction of the surface BZ
calculated within RPA (panel a) and TDLDA (panel b).  The Mg surface
is modeled by a slab of 16 layers as in previous
calculations\cite{Chulkov_L04}, and a vacuum region of 40~\AA. Such
thick slab and vacuum region is necessary to avoid splitting of the
surface plasmon peak due to coupling between the surface plasmons at
the two sides of the slab. The LDA wave functions are calculated on a
uniform grid with a grid spacing of 0.24 \AA~and a $64 \times 64
\times 1$ Monkhorst-Pack $k$-point sampling.  For the response
function calculations we include 200 bands (including 16 occupied
bands) and use a broadening parameter of 0.02 eV.  We use an
anisotropic cutoff energy for the local field
effects\cite{Chulkov_L04}. Since the surface plasmon depends
sensitively on the density profile at the surface where the density
decays exponentially into the vacuum, a cutoff energy of 500 eV is
applied in the $z$-direction. Compared to the RPA results in panel (a),
the inclusions of the LDA exchange-correlation kernel in panel (b) shifts
the peaks down by 0.1 - 0.2 eV.

The energies of these surface plasmons for both the $\bar{\Gamma} - \bar{M}$ and
$\bar{\Gamma} - \bar{K}$ directions are shown in Fig. \ref{fig:Mg_fig} (c). The obtained dispersion relations agree well with previous
calculations\cite{Chulkov_L04}. The  well known negative dispersion at small $\ve
q$ observed for simple metal surfaces are also well reproduced in this work. Compared to
experimental data, the TDLDA energies of the surface plasmons agree within 0.1 eV
for small $\ve q$, while the discrepancy increases to around 0.2 eV  for larger
$\ve q$, which can attributed to the fact that the ALDA kernel is $\ve
q$-independent\cite{Chulkov_L04}.

\begin{figure}[t]
    \centering
    \includegraphics[width=0.95\linewidth]{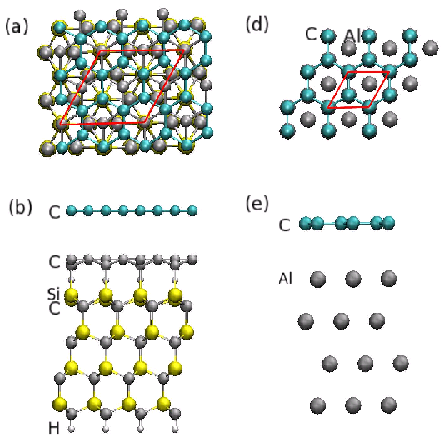}
    \includegraphics[width=1.0\linewidth]{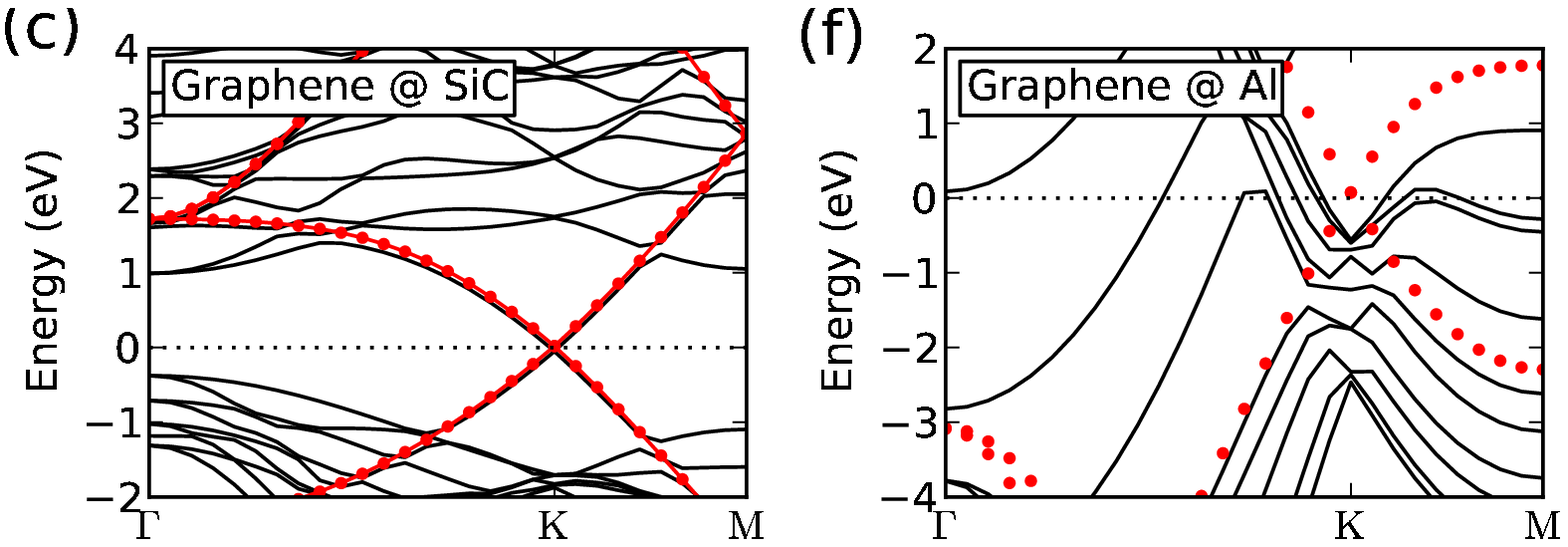}
    \caption{(Color online) Atomic structure of graphene adsorbed on
      SiC(0001) (a+b) and Al(111) (d+e). The lateral unit cells are
      indicated by red lines in the top panels. The LDA band
      structures of the surfaces are shown in the lower panels. Also
      shown is the band structure of free standing graphene (red
      dots). The Fermi level is set to zero.  }
    \label{fig:graphene_substrate_structure}
\end{figure}

\begin{figure}[t]
  \centering
  \includegraphics[width=0.95\linewidth,angle=0]{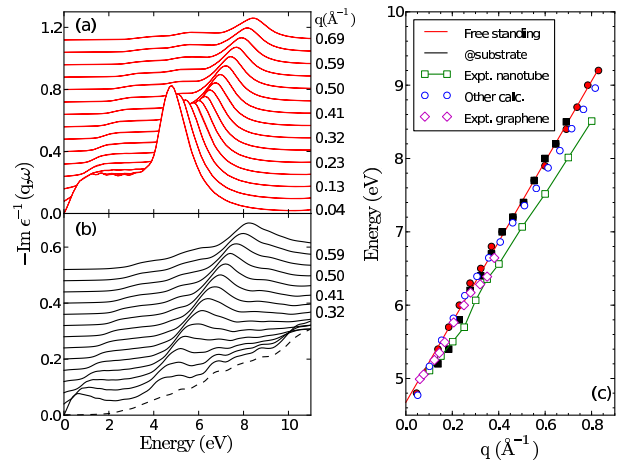}
  \caption[]{ \label{fig:graphene_SiC}(Color online) Loss function of free standing
graphene (a) and
graphene on SiC substrate (b) as a function of q. The loss functions, from bottom to
top (solid lines), correspond to increasing q at an interval of $0.046/$\AA. The
dashed line corresponds to the loss function of the substrate at $q=0.092/$\AA.
(c) Dispersion relations for the $\pi$ plasmons of free standing graphene (red
filled circles)
and graphene on SiC (black filled squares). They are compared with earlier
ab-initio calculation on free standing graphene (blue hollow circles) and
experiments on single wall carbon nanotubes (green hollow
squares)\cite{Reining_L08_2} as well as experiments on graphene / SiC(0001) (purple
hollow diamonds)\cite{Lu_B09}. Lines are added to guide the eye.}
\end{figure}

\subsection{Plasmons in adsorbed graphene}
In this section we investigate the influence of a substrate on the
plasmon excitations in graphene. For a more detailed discussion of
these results we refer the reader to Ref.~\onlinecite{Jun_graphene}. As
representatives for semiconducting and metallic substrates we consider
SiC(0001) and Al(111). Both of these systems bind the graphene
relatively weakly so that hybridization effects are relatively
unimportant. Thus the largest effect of the substrate is expected to
arise from the long range Coulomb interaction between electrons in the
two subsystems.

The atomic structure and band structure of graphene on both substrates
are shown in Fig.~\ref{fig:graphene_substrate_structure}. For
graphene/SiC(0001), the unit cell, indicated by red solid lines in
panel (a), contains $2 \times 2$ graphene and $\sqrt{3} \times
\sqrt{3}$ SiC\cite{SiC_sub1, SiC_sub2}.  As can be seen in panel (b),
two carbon layers are adsorbed on four bi-layers of SiC and the
dangling bonds at the backside of the slab are saturated by hydrogen. The first carbon layer
adsorbs covalently on the SiC surface and is here considered as a part of the substrate. The upper carbon layer
binds weakly to the substrate, in agreement with
experiments\cite{SiC_PES}, with an LDA binding energy per C atom of 0.039 eV, and adsorption distance of 3.56 \AA. As shown in panel (c), linear
conical bands appear within the bandgap of the substrate, resembling
that of free-standing graphene (red dotted line). The Fermi level is
shifted up by 0.05 eV, introducing slight electron doping into
graphene. For the graphene/Al(111) structure we use a $1\times
1$ unit cell with four layers of Al as substrate. Again,
graphene binds weakly to the Al surface with an LDA interplane
distance of 3.36\AA~ and binding energy per C atom of 0.049 eV, in
good agreement with recent van der Waals DFT calculations\cite{Marco_B10}.
As shown in panel (f), the 'Dirac cone' of graphene is shifted $~0.5$ eV below the Fermi level. 
The computational details for calculation of the loss functions can be found in Ref.~\onlinecite{Jun_graphene}.

Fig.~\ref{fig:graphene_SiC} shows the calculated loss function of free standing graphene (a) and graphene on SiC (b).
The freestanding graphene exhibits
a collective mode at around 5 eV, which results from the electronic transitions of the
$\pi \rightarrow \pi^{\ast}$ bands and is referred to as the graphene $\pi$ plasmon.  The dispersion of
the $\pi$ plasmon is shown in Fig. \ref{fig:graphene_SiC}(c). In contrast to its three
dimensional counterpart, graphite, which shows a parabolic dispersion of the $\pi$
plasmons\cite{Rubio_L02}, graphene has a linear plasmon dispersion. The origin of the linear dispersion has been attributed to the role of local field effects\cite{Reining_L08_2}.

Fig. \ref{fig:graphene_SiC} (b) shows the loss function of graphene
adsorbed on the SiC(0001) surface. Compared to the results of the free
standing graphene, the strength of the $\pi$ plasmons are strongly
damped, in particular for small $\ve q$ values.  As $\ve q$ increases,
the strength of the $\pi$ plasmons gradually recovers to that of a
free standing graphene, indicating that the substrate effect becomes
weaker for larger $\ve q$. As shown in Fig. \ref{fig:graphene_SiC}(c),
the substrate has little effect on the energies of the $\pi$ plasmons.
In fact the plasmon dispersion for both free standing and substrate
supported graphene agree well with previous
calculations\cite{Reining_L08_2} as well as experiments on
graphene/SiC\cite{Lu_B09} and carbon nanotubes\cite{Reining_L08_2}.
We have found that the response function, and thus the EELS spectrum,
of the combined graphene/substrate system can be obtained accurately
from the response functions of isolated graphene and substrate
assuming only Coulomb interaction between the two, i.e. neglecting
effects related to hybridization and charge
transfer\cite{Jun_graphene}. This demonstrates that the strong damping
of plasmons results from the non-local screening of the graphene plasmon excitation by the
substrate electrons.

\begin{figure}[t]
  \centering
  \includegraphics[width=1.0\linewidth,angle=0]{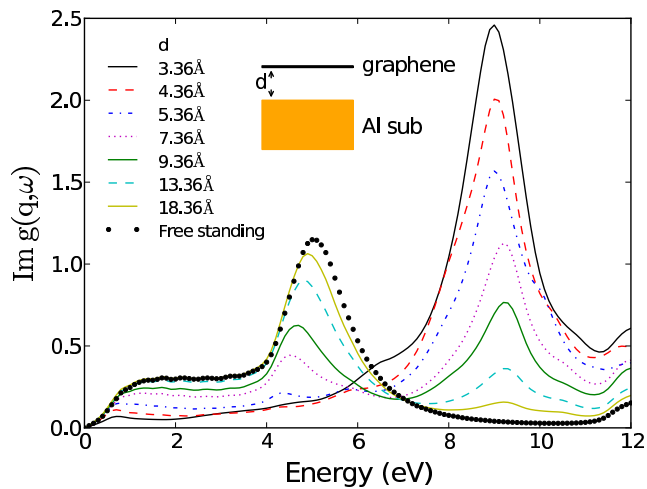}
  \caption[]{ \label{fig:graphene_Al} (Color online) Surface loss functions for
graphene on Al(111) as a
function of the adsorption distance $d$ for $|\ve q|=0.046/$\AA. The surface loss function of
free standing graphene is shown as black dots. Inset: sketch of graphene on
Al substrate.}
\end{figure}

Fig.~\ref{fig:graphene_Al} shows the surface loss function of graphene
on Al(111) for various adsorption distances. In contrast to the
semiconducting SiC substrate, the $\pi$ plasmon at 5 eV is completely
quenched on the metallic Al substrate at the equilibrium distance
$d=3.36$~\AA (full black line). As the graphene is pulled away from
the surface, the $\pi$ plasmon reappears at an energy lower than that
of the free standing graphene. This downshift is due to the coupling to the surface
plasmons of the aluminum substrate at $~9.0$ eV. The graphene $\pi$ is fully
recovered at a distance of around 20~\AA~illustrating the long range nature of the interaction.

\section{Conclusions}
We have implemented the linear density response function in the
adiabatic local density approximation (ALDA) within the real space
projector augmented wave method GPAW, and used it to calculate optical
and dielectric properties of a range of solids, surfaces and
interfaces.  The Kohn-Sham wave functions, from which the response
function is built, can be obtained either on a real space grid or in
terms of localized atomic orbital basis functions. The latter option
reduces the computational requirements for calculating and storing the
often very large number of wave functions required for the
construction of the response function without sacrificing accuracy.
The dielectric constants of a number of bulk semiconductors as well as
the optical absorption spectrum of silicon at the ALDA level was shown
to be in good agreement with previous calculations.  For the surface
plasmons of the Mg(0001) surface we find, in agreement with previous
studies, that the ALDA kernel lowers the plasmon energies by around
0.3 eV realtive to the RPA values and thereby reduces the deviation
from experiments from around 4$\%$ to 1-2$\%$. Very good agreement
with experiments was also found for the plasmon energies of graphene
which were shown to exhibit a linear dispersion with a value of 4.9 eV
in the long wave length limit. The deposition of graphene on a SiC
substrate is shown to have little effects on the plasmon energies but
leads to significant damping of the plasmon resonances. In contrast
deposition on an Al surface completely quenches the graphene plasmons
due to strong non-local electronic screening effects.

\begin{acknowledgments}
The Center for Atomic-scale Materials Design is sponsored
by the Lundbeck Foundation. The Catalysis for Sustainable Energy initiative is
funded by the Danish Ministry of Science, Technology and Innovation. The computational studies were supported as part of the 
Center on Nanostructuring for Efficient Energy Conversion, an Energy 
Frontier Research Center funded by the U.S.
Department of Energy, Office of Science, Office of Basic Energy 
Sciences under Award No. DE-SC0001060.

\end{acknowledgments}


\end{document}